\documentclass[3p,times,twocolumn]{elsarticle}
\pdfoutput=1
\usepackage{ecrc}

\volume{00}

\firstpage{1}

\journalname{Nuclear Physics B Proceedings Supplement}

\runauth{H.~Stenzel}

\jid{nuphbp}

\jnltitlelogo{Nuclear Physics B Proceedings Supplement}

\usepackage{amsmath,amssymb,amstext}
\usepackage{amsthm}
\usepackage{graphicx}

\usepackage{lineno}

\usepackage[figuresright]{rotating}

\def\betastar{\ensuremath{\beta^{\star}}}
\def\sigmatot{\ensuremath{\sigma_{\mathrm{tot}}}} 
\def\rts{ \ensuremath{\sqrt{s}}}
\def\TeV{\ifmmode {\mathrm{\ Te\kern -0.1em V}}\else
                   \textrm{Te\kern -0.1em V}\fi}%
\def\GeV{\ifmmode {\mathrm{\ Ge\kern -0.1em V}}\else
                   \textrm{Ge\kern -0.1em V}\fi}%
\def\MeV{\ifmmode {\mathrm{\ Me\kern -0.1em V}}\else
                   \textrm{Me\kern -0.1em V}\fi}%
\def\keV{\ifmmode {\mathrm{\ ke\kern -0.1em V}}\else
                   \textrm{ke\kern -0.1em V}\fi}%
\def\eV{\ifmmode  {\mathrm{\ e\kern -0.1em V}}\else
                   \textrm{e\kern -0.1em V}\fi}%

\def\stat{\mbox{$\;$(stat.)}}
\def\syst{\mbox{$\;$(syst.)}}

\begin{document}

\begin{frontmatter}



\dochead{}

\title{Measurement of the total cross section from elastic 
scattering in $pp$ collisions at $\rts=7\TeV$  with the ATLAS detector}

\author{Hasko Stenzel\\ On behalf of the ATLAS Collaboration}

\address{Justus-Liebig Universit\"at Gie{\ss}en, II. Physikalisches Institut, Heinrich-Buff Ring 16, D-35392 Giessen, Germany}

\begin{abstract}
In this contribution a measurement of the total $pp$ cross section with the ATLAS detector at the LHC at $\rts=7\TeV$ is presented. 
In a special run with high-\betastar\, beam optics, the differential elastic cross section is measured as a function of 
the Mandelstam momentum transfer variable $t$. The measurement is performed with the ALFA sub-detector of ATLAS. 
Using a fit to the differential elastic cross section in the $|t|$ range from $0.01~\GeV^2$ to $0.1~\GeV^2$ to 
extrapolate to $|t|\rightarrow 0$, the total cross section, $\sigmatot(pp\rightarrow X)$,  
is measured via the optical theorem to be:
\begin{displaymath}
\sigmatot(pp\rightarrow X) =  \mbox{95.35} \; \pm 0.38 \; ({\mbox{stat.}}) \pm 1.25 \; 
({\mbox{exp.}}) \pm 0.37 \; (\mbox{extr.})  \; \mbox{mb} \; \; , 
\end{displaymath}
where the first error is statistical, the second accounts for all experimental systematic uncertainties and the last 
is related to uncertainties in the extrapolation to $|t|\rightarrow 0$. 
In addition, the slope of the elastic cross section at small $|t|$ is determined to be $B = 19.73 \pm 0.14 \; ({\mbox{stat.}}) \pm 0.26  \; ({\mbox{syst.}}) \; \mbox{\GeV}^{-2}$.    
\end{abstract}


\end{frontmatter}


\section{Introduction}
\label{sec:intro}
The total cross section is a fundamental parameter of strong interactions, setting the scale for 
the interaction strength for all processes at a given energy. 
A calculation of the total cross section from first principles, based upon quantum chromodynamics (QCD), 
is not possible, since large distances are involved.  
A measurement of the total cross section can still be performed via elastic scattering by using the 
optical theorem, which relates 
the imaginary part of the forward elastic-scattering amplitude to the total cross section: 
\begin{equation}\label{eq:OpticalTheorem}
\sigmatot \propto \mbox{Im} \, [ f_{\mathrm{el}}\left(t \rightarrow 0\right)] \; ,
\end{equation}
where $f_{\mathrm{el}}(t \rightarrow 0)$ is the elastic-scattering amplitude extrapolated to the forward direction, 
i.e.\ at  $|t|\rightarrow 0$, 
$t$ being the four-momentum transfer. 
In this analysis, a luminosity-dependent method is used to extract the total cross section from 
a fit to the differential elastic cross section according to 
\begin{equation}\label{eq:totxs}
\sigmatot^{2} = \frac{16\pi(\hbar c)^2}{1+\rho^2} \left. \frac{\mathrm{d}\sigma_{\mathrm{el}}}{\mathrm{d}t}\right|_{t \rightarrow 0} \; ,  
\end{equation}
where $\rho$  represents a small correction arising from the ratio of the real to 
imaginary part of the elastic-scattering amplitude in the forward direction and is taken from theory.

More details on the results presented in this paper are given in Ref.~\cite{ConfNote}.  
The quantities measured and reported here have also been measured at the LHC by the 
TOTEM experiment~\cite{totem1,TOTEM_lumindep}. 

\section{Experimental setup}
\label{sec:exp}
\begin{figure}[hb!]
  \centering
  \includegraphics[width=\columnwidth]{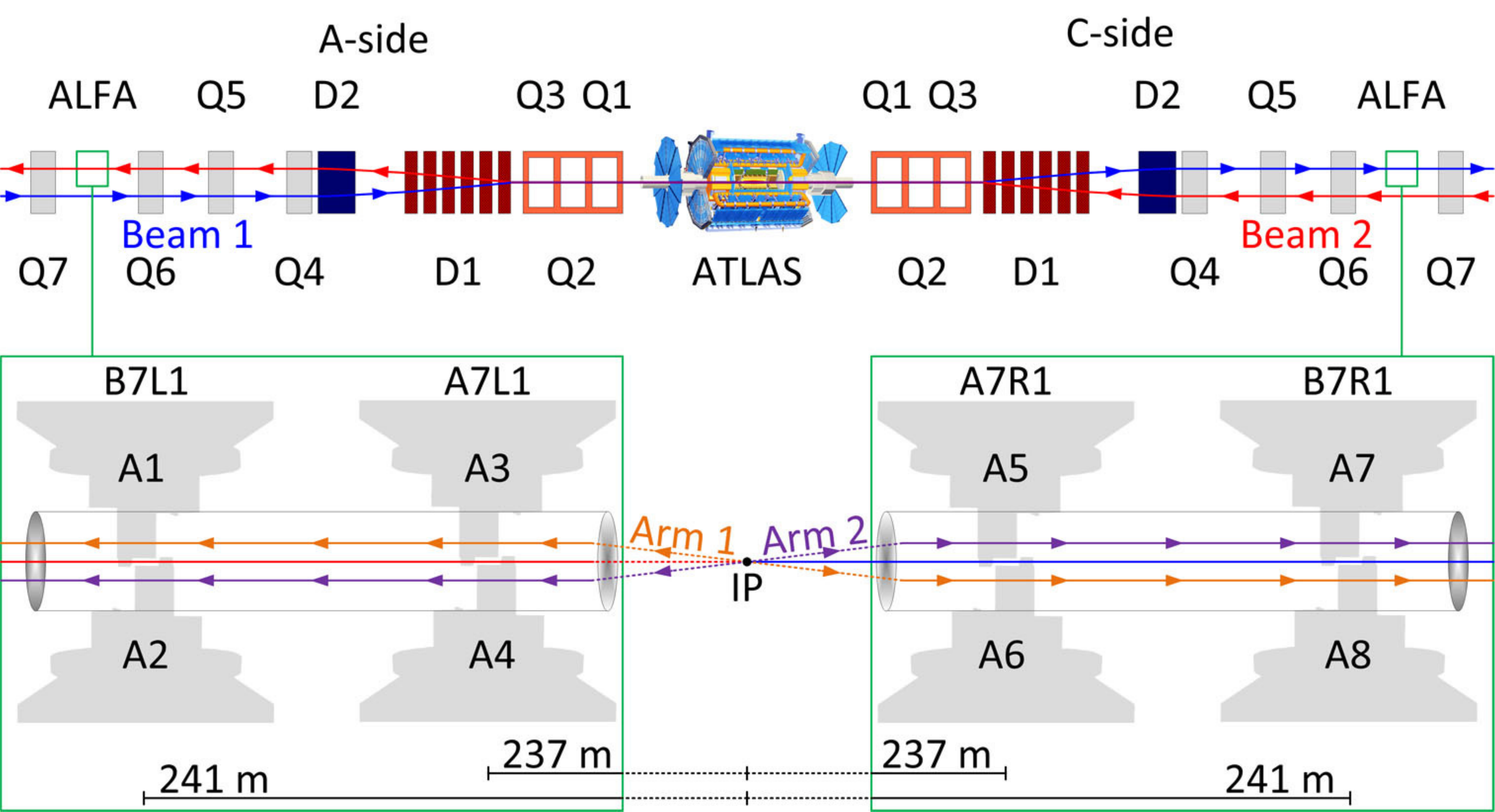} 
  \caption{A sketch of the experimental set-up, not to scale, showing the
   positions of the ALFA Roman Pot stations in the outgoing LHC beams, and the quadrupole
   (Q1--Q6) and dipole (D1--D2) magnets situated between the interaction point and ALFA.
   The ALFA detectors are numbered A1--A8, and are combined into inner stations A7R1 and A7L1, 
   which are closer to the interaction point, and outer stations B7R1 and B7L1. 
   }
  \label{fig:alfa_layout}
\end{figure}

ATLAS is a multi-purpose detector designed to study elementary processes in 
proton--proton interactions at the$\TeV$ energy scale. 
A detailed description of the ATLAS detector can be found in Ref.~\cite{atlas1}. 
Elastic scattering protons are detected with the ALFA (Absolute Luminosity For ATLAS) Roman Pot 
detector system. Two tracking stations are placed on each side of the central ATLAS 
detector at distances of 238 m and 241 m from the interaction point. 
The detectors are housed in so-called Roman Pots (RPs) which can be moved vertically close 
to the circulating proton beams. The RPs are instrumented with main tracking detectors (MDs) 
to determine the track coordinates, with overlap detectors (ODs) to determine the distance 
between the upper and lower detectors, each set of these detectors is supplemented by trigger scintillators.
Each MD consists of 2 times 10 layers of 64 
square scintillating fibres with 0.5~mm side length 
glued on titanium plates. The fibres on the front and back sides of each titanium plate 
are orthogonally arranged at angles of $\pm$45$^\circ$ with respect to the $y$-axis.
The overlap detectors consist of three layers of 30 scintillating fibres per layer measuring 
the vertical coordinate of traversing beam-halo particles or shower fragments.
The station and detector naming scheme is depicted in Fig.~\ref{fig:alfa_layout}.
The stations A7R1 and B7R1 are positioned at $z$ = $-$237.4 m and $z$ = $-$241.5 m respectively 
in the outgoing beam 1 (C side), while the stations A7L1 and B7L1  are situated 
symmetrically in the outgoing beam 2 (A side). 

The data were recorded in a dedicated low-luminosity run using beam optics with a $\betastar$ of 90 m; 
details of the beam optics settings can be found in Refs.~\cite{Note90m,HBLumiday11}. 
For elastic-scattering events, the main pair of colliding bunches was used, which contained around 7$\times$10$^{10}$ protons per bunch.
Very precise positioning of the RPs is mandatory to achieve the 
desired precision on the position measurement of 20--30~$\mu$m in both the horizontal and vertical 
dimensions. 
The first step is a beam-based alignment procedure to determine the position of the RPs with 
respect to the proton beams. In a second step the detector positions  
are directly determined from the elastic-scattering data. 
The alignment procedure is based on the distribution of track positions in
the RP stations in the full elastic-scattering event sample. This distribution forms a
narrow ellipse with its major axis in the vertical ($y$) direction, with an aperture gap
between the upper and lower detectors.
The measured distances between upper and lower MDs and the 
rotation symmetry of scattering angles are used as additional constraints.  
Three parameters are necessary to align each MD: the horizontal and 
vertical positions and the rotation angle around the beam axis. 
The horizontal detector positions and the rotation angles are determined from a fit of a straight 
line to a profile histogram of the narrow track patterns in the upper and lower MDs.     
The uncertainties are 
1--2~$\mu$m for the horizontal coordinate and 0.5~mrad for the angles.
For the vertical detector positioning, the essential input is the distance between the upper 
and lower MDs. The achieved precision on the vertical alignment is about 80 $\mu$m.

To trigger on elastic-scattering events, two main triggers were used. 
The triggers 
required a coincidence of the main detector trigger scintillators between either
of the two upper (lower) detectors on side A and either of the two lower (upper) detectors on side C. 
The trigger efficiency for elastic-scattering events was determined from a data stream in which all events with a hit in 
any one of the ALFA trigger counters were recorded. In the geometrical acceptance of the detectors, the efficiency of the trigger used to record 
elastic-scattering events is $99.96 \pm 0.01\%$.

\section{Measurement principle and beam optics}
The data were recorded with a beam optics of high   
$\betastar$ of 90 m resulting in a small divergence and 
providing parallel-to-point focusing in the vertical plane.  
In parallel-to-point beam optics  
the betatron oscillation has a phase advance $\Psi$ of $90^\circ$ between the interaction point 
and the RPs, such that all particles scattered 
at the same angle are focused at the same position at the detector, independent of their production vertex position. 
The four-momentum transfer $t$ is calculated from the scattering angle $\theta^\star$; 
in elastic scattering at high energies this is given by:
\begin{equation}
\label{eq:t-basic}
 -t = \left(\theta^\star \times p \right)^2 \,\,,
\end{equation}
where $p$ is the nominal beam momentum of the LHC of $3.5 \TeV$ and 
$\theta^\star$ is measured from the proton tra\-jectories in ALFA.  
The trajectory ($w(z)$, $\theta_w(z)$), where $w=x,y$ is the transverse position with 
respect to the nominal orbit at a distance $z$ from 
the interaction point and $\theta_w$ is the angle between $w$ and $z$, is 
given by the transport matrix {\bf M} and the coordinates at 
the interaction point ($w^\star$, $\theta_w^\star$):

\begin{equation}\label{eq:transport}
\left(\begin{array}{c}
w(z) \\
\theta_w(z) \\
\end{array}\right) =
{\bf M} 
\left(\begin{array}{c}
w^\star \\
\theta_w^\star \\
\end{array}\right) =
\left(\begin{array}{cc}
M_{11} & M_{12} \\
M_{21} & M_{22} 
\end{array}\right)
\left(\begin{array}{c}
w^\star \\
\theta_w^\star \\
\end{array}\right) \;  \; ,
\end{equation}
where the elements of the transport matrix can be calculated from the 
optical function $\beta$ and its derivative with respect to $z$ and $\Psi$. 
The $t$-reconstruction is based upon the track positions and certain transport matrix elements. 
The ALFA detector was designed to use the ``subtraction'' method to calculate the scattering angle:  
\begin{equation}
\label{eq:subtraction}
\theta^\star_w = \frac{w_\mathrm{A} - w_\mathrm{C}}{M_{12,\mathrm{A}} + M_{12,\mathrm{C}}} \; \; ,
\end{equation}
exploiting the 
fact that for elastic scattering the particles are back-to-back, the scattering angle at the A- and C-side 
are the same in magnitude and opposite in sign, and the protons originate from the same vertex. 
Three different alternative methods ``local subtraction'', ``lattice'' and ``local angle''~\cite{ConfNote} using 
a combination of the track position and angle between inner and outer stations and different sets 
of matrix elements are used to derive constraints on the beam optics and as a cross-check for the nominal 
subtraction method. 
For all methods $t$ is calculated from the scattering angles as follows:
\begin{eqnarray}
- t & = &\left((\theta^\star_{x})^2 + (\theta^\star_{y})^2  \right)p^2   \; \; .
\end{eqnarray}

\begin{figure}[h!]
  \centering
  \includegraphics[width=\columnwidth]{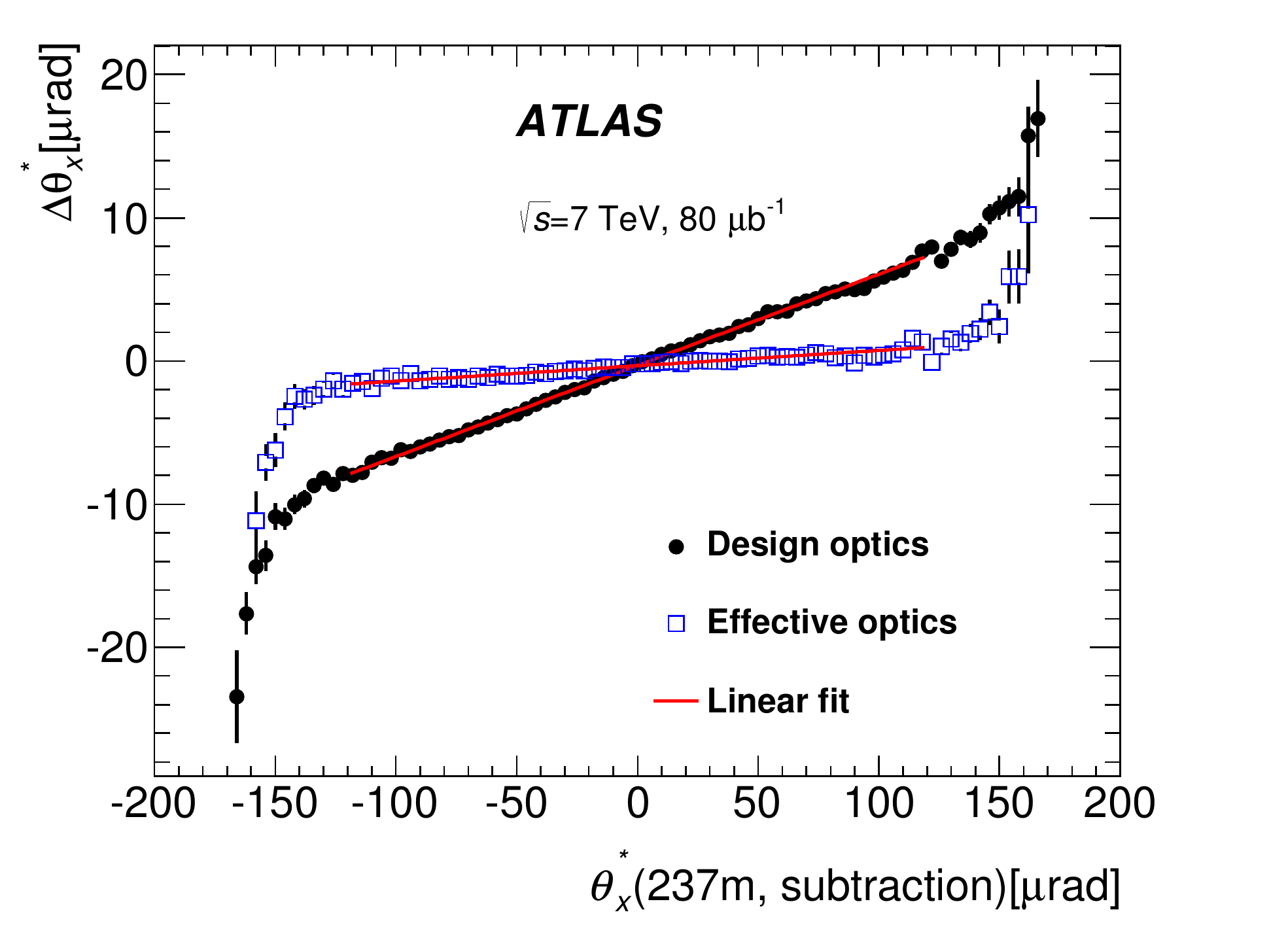}
  \caption{The difference in reconstructed scattering angle $\Delta\theta^\star_x$ between the subtraction and local angle 
  methods as a function 
  of the scattering angle from subtraction method for the inner detectors. The line represents the result of a linear fit. 
  Values obtained using the tuned effective optics are also shown for comparison.} 
  \label{fig:per_x}
\end{figure}

The precision of the $t$-reconstruction depends on knowledge of the elements of the transport matrix. 
From the design of the 90~m beam optics along with the alignment parameters of the magnets, the magnet currents and the 
field calibrations, all transport matrix elements can be calculated. 
This initial set of matrix elements is referred to as ``design optics''. 
Small corrections, allowed within the range of the systematic uncertainties, 
need to be applied to the design optics for the measurement of $\sigmatot$.
Constraints on beam optics parameters 
are derived from the ALFA data, exploiting the fact that the reconstructed scattering angle 
must be the same for different reconstruction methods using different transport matrix elements.   
Two classes of constraints are distinguished:
\begin{itemize}
\item Correlations between positions or angles measured either at the A-side and C-side 
or at inner and outer stations of ALFA. These are used to infer the ratio of matrix elements in the beam transport matrix. 
The resulting constraints are independent of any optics input.  
\item Correlations between the reconstructed scattering angles. These are calculated using different methods 
to derive further constraints on matrix elements as scaling factors. These factors 
indicate the amount of scaling needed to be applied to a given matrix element ratio in order 
to equalize the measurement of the scattering angle. These constraints depend on the given optics model.
The design beam optics with quadrupole currents measured during the run is used as reference to calculate the constraints. 
\end{itemize}  
The best example is the comparison
of the scattering angle in the horizontal plane reconstructed with the subtraction method, 
which is based on the position and $M_{12,x}$,  
and with the local angle method, which is based on the local angle and $M_{22,x}$. 
The scaling factor for the matrix element ratio $M_{12}/M_{22}$ is derived from the slope 
of the difference of the scattering angle obtained with the two methods as a function of the scattering 
angle determined with the subtraction method, as shown in Fig.~\ref{fig:per_x}. In total 14 constraints are derived from ALFA data. 

The constraints are combined in a fit used to determine the beam optics. 
The free parameters of the fit are the quadrupole strengths in both beams.
The chosen configuration, 
called the effective optics, is one solution among many. This solution is obtained by allowing only the inner 
triplet magnets Q1 and Q3 to vary coherently from their nominal strength. Q1 and Q3 were manufactured at a different site from 
the other quadrupoles, and relative calibration differences are possible. 
The fit resulted in an 
offset of approximately 0.3\% for the strength of Q1 and Q3, with a difference of about $10\%$ between 
the two beams and the uncertainty from the fit is also of $10\%$.  
The $\chi^{2}$ of the fit includes the systematic uncertainties of the constraints and is of good quality 
with $\chi^{2}/N_{\mbox{dof}}=1.1$. 
This effective optics is used for the total cross-section measurement.

\section{Data Analysis}
\label{sec:data_analysis}
Events are required to pass the trigger conditions for elastic-scattering events, 
and have a reconstructed track in all four detectors of the
arm which fired the trigger. 
Events with additional tracks in  
detectors of the other arm arise from the overlap of halo protons with 
elastic-scattering protons and are retained. 
Further geometrical cuts on the left-right acollinearity are applied, exploiting the back-to-back topology of 
elastic-scattering events. The position difference between the left and the right sides is required to be within 3.5$\sigma$ 
of its resolution determined from simulation. An efficient cut against non-elastic background is obtained from 
the correlation of the local angle between two stations and the position in the horizontal plane. 
Finally, fiducial cuts to ensure a good containment inside the detection area are applied to the 
vertical coordinate. It is required to be at least 60~$\mu$m from the edge of the detector nearer the beam, where the full detection efficiency 
is reached. At large vertical distance,  the vertical coordinate must be at least 1 mm away from the shadow 
of the beam screen, a protection 
element of the quadrupoles, in order to minimize the impact from showers generated in the beam screen.
At the end of the selection procedure 805,428 events survive all cuts. 

A small fraction of background events is expected to survive the elastic event selection cuts.  
The background events peak  
at small values of $x$ and $y$ and thus constitute an irreducible background at small $t$. 
The background predominantly originates from accidental coincidences of beam-halo particles, but  
single diffractive protons in coincidence with a halo proton at the opposite side may also contribute. 
The irreducible background fraction is estimated by counting events in the ``anti-golden'' topology 
with two tracks in both upper or both lower detectors~\cite{ConfNote}, which also allows constructing 
the $t$-spectrum and subtracting it from the spectrum of the selected sample. The background 
contamination is about 0.5$\%$ with large systematic uncertainties of 50--80\%, which are dominated 
by the background normalization uncertainty.  

Elastic-scattering events inside the acceptance region are expected to have a proton track in each of the four detectors of 
the corresponding spectrometer arm. Losses of elastic events occur in the case of interactions of the protons with the 
stations or detectors, which result in too large fibre hit multiplicities and a failure of the track reconstruction algorithm. 
The event reconstruction efficiency is used to correct for these losses and 
its determination is based on a tag-and-probe approach using a data-driven method. 
Elastic events not fully reconstructed and thus not used for the analysis are grouped into several reconstruction cases, 
according to the number of detectors with a reconstructed track out of the four detectors ideally present. 
The reconstruction efficiency is given by: 
\begin{equation}\label{eq:eff2}
\varepsilon_{\text{rec}} = \frac{N_{4/4}}{N_{4/4}+N_{3/4}+N_{2/4}+N_{1+1/4}+N_{1/4}+N_{0/4}},
\end{equation}
where $N_{k/4}$ is the number of events with $k$ detectors with at least one reconstructed track in a spectrometer arm 
and $N_{1+1/4}$ is the number of events with a track reconstructed in only one detector at each side. 
The event reconstruction efficiencies in arm 1 and arm 2 are determined to be 
$\varepsilon_{\mathrm{rec,1}} = 0.8974 \pm 0.0004 \stat \pm 0.0061 \syst$ and $\varepsilon_{\mathrm{rec,2}} = 0.8800 \pm 0.0005 \stat \pm 0.0092 \syst$ respectively. 

ATLAS exploits several detectors and algorithms to determine the luminosity and evaluate the related systematic uncertainty~\cite{LumiPaper2011}. 
The absolute luminosity scale of each algorithm was calibrated~\cite{LumiPaper2011} by the van der Meer ({\it vdM}) method~\cite{svdm}
in an intermediate luminosity regime ($L\,\sim 5 \times 10^{30}$\,cm$^{-2}$\,s$^{-1}$). 

The conditions in the low-luminosity run analysed here are very different from those in high-luminosity runs. 
The instantaneous luminosity is about six orders of magnitude lower ($L\,\sim 5 \times 10^{27}$\,cm$^{-2}$\,s$^{-1}$), 
the beam--gas contribution, normally negligible, can become competitive with the collision rate, but 
the background due to slowly decaying, collision-induced radiation 
(often called ``afterglow'' ~\cite{LumiPaper2011}) becomes conversely less important. 
The luminosity is determined using the BCM (beam conditions monitor), as in Ref.~\cite{LumiPaper2011}, and other detectors 
and algorithms are used to assess the systematic uncertainty. 
The integrated luminosity for the selected running period is:
\begin{equation*}
L_{\mathrm{int}}=78.7 \pm 0.1 \,(\mathrm{stat.}) \pm 1.9 \,(\mathrm{syst.})  ~\mu \mathrm{b}^{-1} \; ,\label{LumiWithSys}
\end{equation*}
and the total systematic uncertainty amounts to 2.3\%, which comprises 
the scale uncertainty, the overall calibration-transfer uncertainty and the background uncertainty. 

Simulations for the calculation of the acceptance corrections are carried out with PYTHIA8 \cite{PYTHIA,PYTHIA6}. 
Generated particles are transported from the interaction point to the RPs using either the transport 
matrix Eq.~\eqref{eq:transport} or the MadX~\cite{madx} beam optics calculation package. A fast parameterization of 
the detector response is used for the detector simulation with the detector resolution tuned to the measured resolution.
The measured $t$-spectrum in each arm, after background subtraction, is corrected for migration 
effects using an iterative, dynamically stabilised unfolding method~\cite{IDS}. 
A data-driven closure test is used to evaluate any bias in the unfolded data spectrum shape due 
to mis-modelling of the reconstruction-level spectrum shape in 
the simulation. 

\section{Results}
In order to calculate the differential elastic cross section, several corrections are applied. 
The corrections are done individually per detector arm and the corrected spectra from the two arms are combined. 
In a given bin $t_i$, the cross section is calculated according to the following formula:
\begin{equation}\label{eq:cross-section}
\frac{\mathrm{d}\sigma}{\mathrm{d}t_i} = \frac{1}{\Delta t_i}\times \frac{{\cal M}^{-1}[N_i - B_i]}{A_i  \times \epsilon^{\mathrm{reco}} \times \epsilon^{\mathrm{trig}} \times \epsilon^{\mathrm{DAQ}}  \times L_{\mathrm{int}} }\; \; , 
\end{equation}
where $\Delta t_i$ is the bin width, ${\cal M}^{-1}$ represents the unfolding procedure applied to the 
background-subtracted number of events $N_i - B_i$, $A_i$ is the acceptance,  
$\epsilon^{\mathrm{reco}}$ is the event reconstruction efficiency, $\epsilon^{\mathrm{trig}}$ is the trigger efficiency, 
$\epsilon^{\mathrm{DAQ}}$ is the dead-time correction and $L_{\mathrm{int}}$ is the integrated luminosity used for this
analysis. 
The following uncertainties are propagated to the differential elastic cross section:
\begin{itemize}
\item The amount of background is varied by the difference between the yields obtained from 
different methods and different background shapes obtained from variations of the anti-golden method.
\item The impact of alignment uncertainties is estimated from different sets of the alignment parameter values.  
\item The uncertainties related to the effective optics are obtained by varying the optics constraints, 
changing the strength and alignment of the quadrupoles, propagating the fit 
uncertainties to the resulting optics and varying beam transport parameters.
\item The nuclear slope used in the simulation is varied conservatively by $\pm1$ \GeV$^{-2}$ around 19.5 \GeV$^{-2}$ .  
\item Variations of the model for the detector parameterization in the simulation.
\item Variations of the emittance used to calculate the angular divergence in the simulation.
\item The event reconstruction efficiency is varied by its uncertainty.
\item Variations of the track reconstruction parameters.
\item The intrinsic unfolding uncertainty is determined from the data-driven closure test. 
\item The impact of a residual beam crossing angle in the horizontal plane of $\pm10\, \mu$rad is taken into account. 
This variation is derived from the precision of the beam position monitors.
\item The nominal beam energy used in the $t$-reconstruction is changed by $0.65\%$ \cite{Wenninger}. 
\item The luminosity uncertainty of 2.3$\%$ is propagated to the cross section.
\end{itemize}

\begin{figure}[h!]
  \centering
  \includegraphics[width=\columnwidth]{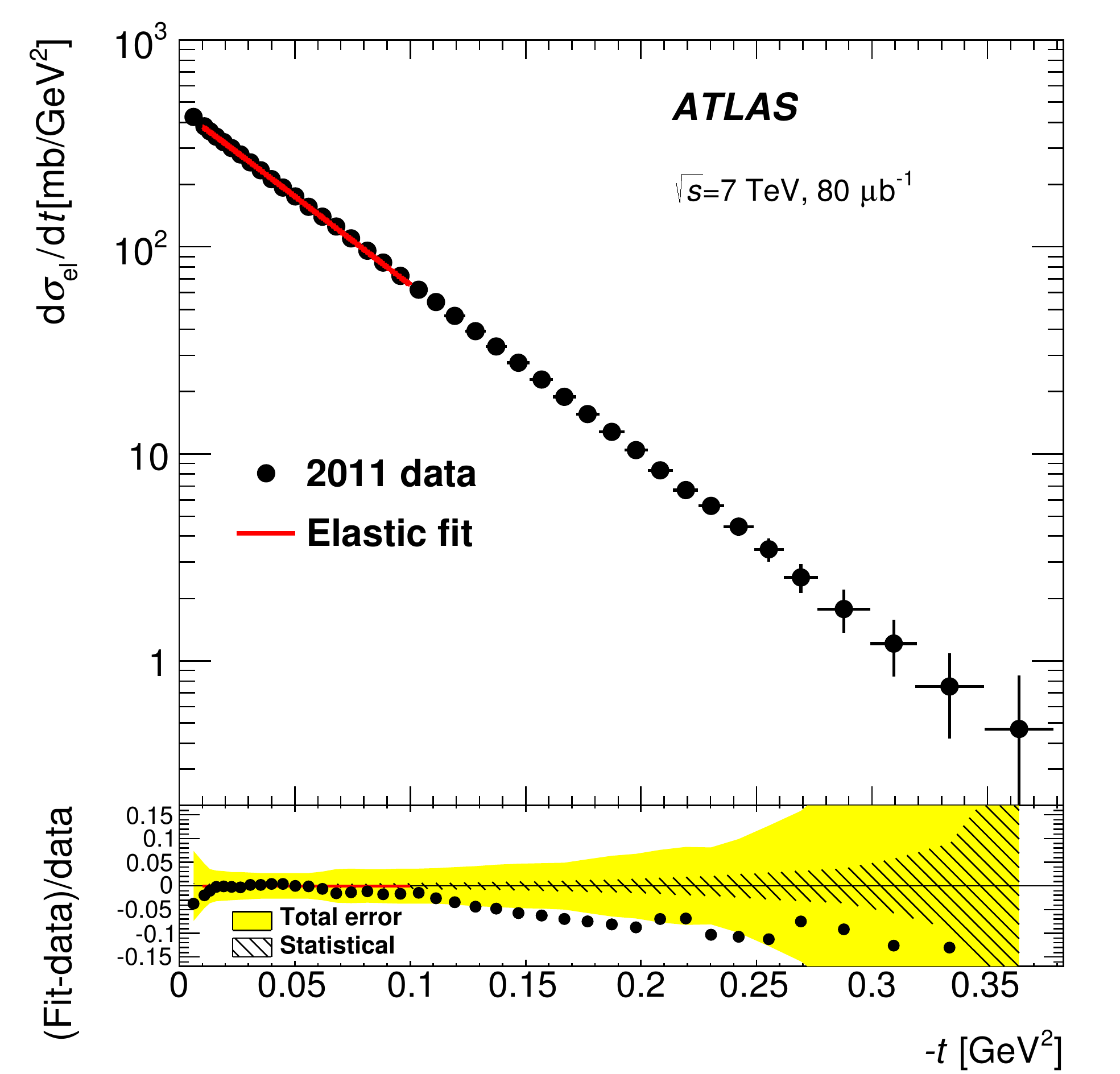}
  \caption{A fit of the theoretical prediction with $\sigmatot$ and $B$ as free parameters to the
 differential elastic cross section reconstructed with the subtraction method.
In the lower panel the points represent the normalized difference between fit and data, 
 the yellow area represents the total experimental uncertainty and the hatched area the statistical component. 
 The red line indicates the fit range.} 
  \label{fig:fit_subtraction}
\end{figure}

The total cross section and the slope parameter $B$ are obtained from a fit of the theoretical spectrum 
\begin{eqnarray}\label{eq:tgen}
\frac{\mathrm{d}\sigma}{\mathrm{d}t} & = & \frac{4\pi\alpha^2(\hbar c)^2}{\mid t \mid^2} \times G^4(t) \\ \nonumber
 & - &  \; \; \sigmatot \times \frac{\alpha G^2(t)}{|t|}\left[\sin\left(\alpha\phi(t)\right) + \rho \cos\left(\alpha\phi(t)\right) \right] \\ \nonumber 
 & \times & \exp{\frac{-B\mid t \mid}{2}} \\ \nonumber
 & + &  \; \; \sigmatot^2 \frac{1+\rho^2}{16\pi(\hbar c)^2} \times \exp\left({-B\mid t \mid}\right) \; \; ,
\end{eqnarray}
where $G$ is the electric form factor of the proton \cite{Cahn}, 
$\phi$ is the Coulomb phase \cite{Bethe,WestAndYennie} and 
the value of $\rho=0.140\pm0.008$ is taken from Ref.~\cite{compete}. 
Both the statistical and systematic uncertainties as well as their correlations are taken into account in the fit, 
shown in Fig.~\ref{fig:fit_subtraction}, which 
uses a profile minimization procedure~\cite{profile}. 
Additional uncertainties arise from the extrapolation $|t|\rightarrow 0$. 
These are estimated from a variation of the upper end of the fit range 
from $-t=0.1 \GeV^2$ to $-t=0.15 \GeV^2$~\cite{per_theory}. 
The upper fit-range edge is also decreased by the same number of bins (six) to $|t|=0.059 \GeV^2$ and the symmetrized 
change is adopted as a systematic uncertainty. Further 
uncertainties comprise the variation of $\rho$ by its uncertainty and variations of the electric form factor 
and Coulomb phase \cite{ConfNote}. The results including uncertainties are summarized in Table~\ref{tab:results}.
\begin{table}
  \begin{center}
    \begin{tabular}{lcc}
      \hline \hline        
	             & $\sigmatot$ [mb] & $B \; [\GeV^{-2}$] \\ \hline 
Central result       & 95.35 & 19.73 \\ 
Statistical error    & 0.38  & 0.14  \\ 
Experimental error   & 1.25  & 0.19  \\ 
Extrapolation error  & 0.37  & 0.17  \\ \hline
Total  error         & 1.36  & 0.35  \\ \hline 
    \end{tabular}
  \caption{Results for the total cross section and nuclear slope.}
  \label{tab:results}
  \end{center}
\end{table}
Several cross-checks of the analysis were carried out, including the investigation of  
the four different $t$-reconstruction algorithms, replacing the profile fit with a simple 
$\chi^2$-minimization using statistical uncertainties only, analyzing the total cross section 
separately in each arm of ALFA, replacing the unfolding procedure of the data by a method where all 
corrections are applied to the theoretical prediction, which is then fit to raw data and finally 
the full sample was split into several time-ordered sub-samples to investigate a potential time dependence. 
All cross-checks gave consistent results. Additionally, different theoretical predictions including 
possible non-exponential terms in the nuclear amplitude were used to extract the total cross section 
and the scatter of results was found consistent with the extrapolation uncertainty~\cite{ConfNote}.

From the fitted parameterization of the differential elastic cross section several additional quantities can be 
derived. The total elastic cross section is obtained from the nuclear scattering term, whereas the Coulomb and 
interference terms are not taken into account:
\begin{eqnarray}
\sigma_{\mathrm{el}} & = & \int_{t=0}^{t=\infty}\sigmatot^2 \frac{1+\rho^2}{16\pi(\hbar c)^2} \, \exp\left({-B\mid t \mid}\right)\mathrm{d}t \\ \nonumber
& = & \frac{\sigmatot^2}{B} \; \frac{1+\rho^2}{16\pi(\hbar c)^2} \; .
\end{eqnarray}  
The differential cross section at the optical point, $|t|\rightarrow 0$, derived from the total cross-section 
fit, is $\mathrm{d}\sigma_{\mathrm{el}}/\mathrm{d}t|_{t\rightarrow 0} = 474 \pm 4 \; (\mbox{stat.}) \pm 13 \;(\mbox{syst.})$ mb/\GeV$^2$, 
where the systematic uncertainty includes all experimental and extrapolation uncertainties.   
Integrating the parameterized form of the differential cross section over the full $t$-range 
yields the total elastic cross section:
\begin{displaymath}
\sigma_{\mathrm{el}} = 24.00 \pm 0.19 \; (\mbox{stat.}) \pm 0.57 \; (\mbox{syst.}) \; \mbox{mb}.  
\end{displaymath}
The measured integrated elastic cross section in the fiducial range from $-t=0.0025$ \GeV$^2$ to $-t=0.38$ \GeV$^2$
corresponds to $90\%$ of the total elastic cross section:
\begin{displaymath}
\sigma_{\mathrm{el}}^{\mathrm{observed}} = 21.66 \pm 0.02 \;(\mbox{stat.}) \pm 0.58 \; (\mbox{syst.}) \; \mbox{mb}.
\end{displaymath} 
The total elastic cross section is used to determine the total inelastic cross section 
by subtraction from the total cross section. The resulting value is:
\begin{displaymath}
\sigma_{\mathrm{inel}} = 71.34 \pm 0.36 \; (\mbox{stat.}) \pm 0.83 \; (\mbox{syst.}) \; \mbox{mb}.
\end{displaymath}

\section{Discussion}
The result for the total hadronic cross section presented here, $\sigmatot=95.35 \pm 1.36$ mb, can be compared to the value measured by TOTEM 
in the same LHC fill using a luminosity-dependent analysis, 
$\sigmatot=98.6 \pm 2.2$~mb \cite{TOTEM_second}. Assuming the uncertainties are uncorrelated, the difference 
between the ATLAS and TOTEM values corresponds to 1.3$\sigma$. The uncertainty on the TOTEM result is dominated by the 
luminosity uncertainty of $\pm 4 \%$, while the measurement reported here profits from a smaller luminosity uncertainty 
of only $\pm2.3\%$.
The value of the nuclear slope parameter $B=19.73 \pm 0.29$ \GeV$^{-2}$ is 
in good agreement with the TOTEM measurement of  $ 19.89 \pm 0.27$ \GeV$^{-2}$ \cite{TOTEM_second}. 
These large values of the $B$-parameter confirm that elastically scattered protons continue
to be confined to a gradually narrowing cone. 
The elastic cross section is measured to be $24.0\pm 0.6$ mb. This is in agreement  with the TOTEM result of $25.4\pm 1.1$ mb within 1.1$\sigma$.
The ratio of the elastic cross section to the total cross section is often taken as a measure of the opacity of the proton.
Measurements shed light on whether the black disc limit of a ratio of 0.5 is being approached. 
The TOTEM value is 
$\sigma_{\mathrm{el}}/\sigmatot =0.257 \pm0.005$ \cite{TOTEM_lumindep,TOTEM_inel}, while the measurement reported here gives
$\sigma_{\mathrm{el}}/\sigmatot=0.252\pm0.004$. All derived measurements depend on $\sigmatot$ and $B$ and are therefore highly correlated.

\section{Conclusion}
In this paper a measurement of the elastic $pp$ cross section and the determination of the total cross section  using the optical theorem at $\sqrt{s}=7 \TeV$ by the ATLAS experiment 
at the LHC with the ALFA sub-detector is presented. The data were recorded in 2011 during a special 
run with high-$\betastar$ optics, where an integrated luminosity of 
$80$ $\mu$b$^{-1}$ was accumulated. The analysis uses data-driven methods to determine relevant 
beam optics parameters, event reconstruction efficiency and to tune the simulation. A key element of 
this analysis is the determination of the effective beam optics, which takes into account measurements 
from ALFA that are sensitive to ratios of transport matrix elements and calibration uncertainties of the quadrupoles. 
A detailed evaluation of the associated systematic uncertainties includes  the comparison of 
different $t$-reconstruction methods that are sensitive to different transport matrix elements.  
A dedicated effort was made to determine the absolute luminosity for this run while
taking into account the  special conditions with a very low number of interactions per bunch crossing. 
From a fit to the differential elastic cross section, using the optical theorem, the total cross section is determined to be:
\begin{eqnarray*}
\sigmatot(pp\rightarrow X) & = &  \mbox{95.35} \; \pm \mbox{1.36} \; \mbox{mb} \; ,
\end{eqnarray*}
where the total error includes statistical, experimental systematic and extrapolation uncertainties.  
The experimental systematic uncertainty is dominated by the uncertainty on the luminosity and on the 
nominal beam energy. 
In addition, the slope of the elastic differential cross section at small $t$ was determined to be:
\begin{equation*}
B = \mbox{19.73} \; \pm \mbox{0.35} \;  \mbox{GeV}^{-2} \; . 
\end{equation*}

More elastic data were recorded at $\rts=8\TeV$ with a high-$\betastar$ optics of 90 m and 1 km, which will 
allow probing the Coulomb-nuclear interference regime at yet smaller values of $t$. Additionally, during 
the shut-down of the LHC a substantial consolidation program was carried out to improve the performance of 
the ALFA detector in run 2 of the LHC. 

\nocite{*}
\bibliographystyle{elsarticle-num}
\bibliography{elastics7}

\begin{thebibliography}{10}
\expandafter\ifx\csname url\endcsname\relax
  \def\url#1{\texttt{#1}}\fi
\expandafter\ifx\csname urlprefix\endcsname\relax\def\urlprefix{URL }\fi
\expandafter\ifx\csname href\endcsname\relax
  \def\href#1#2{#2} \def\path#1{#1}\fi

\bibitem{ConfNote}
{ATLAS Collaboration}, {Measurement of the total cross section from elastic
  scattering in $pp$ collisions at $\rts=7\TeV$ with the ATLAS detector },
  submitted to Nucl. Phys. B.
\newblock \href {http://arxiv.org/abs/1408.5778} {\path{arXiv:1408.5778}}.

\bibitem{totem1}
{G.~Antchev, et al. (TOTEM Collaboration)}, {Double diffractive cross-section
  measurement in the forward region at LHC}, Phys. Rev. Lett. 111 (2013)
  012001.
\newblock \href {http://arxiv.org/abs/1308.6722} {\path{arXiv:1308.6722}}.

\bibitem{TOTEM_lumindep}
{G.~Antchev, et al. (TOTEM Collaboration)}, {Luminosity-independent
  measurements of total, elastic and inelastic cross-sections at $\sqrt{s}=7$
  TeV}, Europhys. Lett. 101 (2013) 21004.

\bibitem{atlas1}
{ATLAS Collaboration}, {The ATLAS Experiment at the CERN Large Hadron
  Collider}, JINST 3 (2008) S08003.

\bibitem{Note90m}
H.~Burkhardt, et~al., {90m Optics Studies and Operation in the LHC}, Conf.
  Proc. C1205201 (2012) 130.

\bibitem{HBLumiday11}
H.~Burkhardt, S.~Cavalier, P.~Puzo, A.~Peskov, {90 m Beta* Optics for
  ATLAS/ALFA}, Conf. Proc. C110904 (2011) 1798.

\bibitem{LumiPaper2011}
{ATLAS Collaboration}, {Improved luminosity determination in pp collisions at
  $\sqrt{s} = 7$ TeV using the ATLAS detector at the LHC}, Eur. Phys. J. C 73
  (2013) 2518.
\newblock \href {http://arxiv.org/abs/1302.4393} {\path{arXiv:1302.4393}}.

\bibitem{svdm}
S.~van~der Meer, {Calibration of the effective beam height in the ISR},
  ISR-PO-68-31 (1968) http://cds.cern.ch/record/296752.

\bibitem{PYTHIA}
T.~Sj\"{o}strand, S.~Mrenna, P.~Skands, {A Brief Introduction to PYTHIA 8.1},
  Comput. Phys. Commun. 178 (2008) 852.
\newblock \href {http://arxiv.org/abs/0710.3820} {\path{arXiv:0710.3820}}.

\bibitem{PYTHIA6}
T.~Sj\"{o}strand, S.~Mrenna, P.~Skands, {PYTHIA 6.4 physics and manual}, JHEP
  0605 (2006) 026.
\newblock \href {http://arxiv.org/abs/hep-ph/0603175}
  {\path{arXiv:hep-ph/0603175}}.

\bibitem{madx}
{CERN Accelerator Beam Physics Group}, Mad - methodical accelerator design,
  {http://mad.web.cern.ch/mad/} (2014).

\bibitem{IDS}
B.~Malaescu, {An Iterative, Dynamically Stabilized (IDS) Method of Data
  Unfolding. }\href {http://arxiv.org/abs/1106.3107} {\path{arXiv:1106.3107}}.

\bibitem{Wenninger}
J.~Wenninger, {Energy Calibration of the LHC Beams at 4~TeV},
  {CERN-ATS-2013-040, http://cds.cern.ch/record/1546734} (2013).

\bibitem{Cahn}
R.~N. Cahn, Coulombic-hadronic interference in an eikonal model, Z. Phys. C 15
  (1982) 253.

\bibitem{Bethe}
H.~A. Bethe, {Scattering and polarization of protons by nuclei}, Ann. Phys. 3
  (1958) 190.

\bibitem{WestAndYennie}
G.~B. West, D.~R. Yennie, Coulomb interference in high-energy scattering, Phys.
  Rev. 172 (1968) 1413.

\bibitem{compete}
J.~Cudell, et~al., {Benchmarks for the Forward Observables at RHIC, the
  Tevatron Run II and the LHC}, Phys. Rev. Lett. 89 (2002) 201801.
\newblock \href {http://arxiv.org/abs/hep-ph/0206172}
  {\path{arXiv:hep-ph/0206172}}.

\bibitem{profile}
V.~Blobel, {Some Comments on $\chi^2$ Minimization Applications }, {eConf} {C
  030908} (2003) {MOET002.
  http://www.slac.stanford.edu/econf/C030908/proceedings.html}.

\bibitem{per_theory}
V.~A. Khoze, A.~D. Martin, M.~G. Ryskin, {Soft diffraction and the elastic
  slope at Tevatron and LHC energies: A multi-Pomeron approach}, Eur. Phys. J.
  C 18 (2000) 167.
\newblock \href {http://arxiv.org/abs/hep-ph/0007359}
  {\path{arXiv:hep-ph/0007359}}.

\bibitem{TOTEM_second}
{G.~Antchev, et al. (TOTEM Collaboration)}, {Measurement of proton-proton
  elastic scattering and total cross-section at $\sqrt{s}=~7$ TeV}, Europhys.
  Lett. 101 (2013) 21002.

\bibitem{TOTEM_inel}
{G.~Antchev, et al. (TOTEM Collaboration)}, {Measurement of proton-proton
  inelastic scattering cross-section at $\sqrt{s}=~7$ TeV}, Europhys. Lett. 101
  (2013) 21003.

\end{thebibliography}







\end{document}